\documentstyle[11pt]{article}
\catcode`\@=11
\begin{document} \openup6pt

\title{ Holographic Dark Energy Model \\  with
Modified Generalized Chaplygin Gas }

\author{B. C. Paul\thanks{Electronic mail : bcpaul@iucaa.ernet.in}  \\
    Physics Department, North Bengal University, \\
Siliguri, Dist. : Darjeeling, Pin : 734 013, West Bengal, India \\
P. Thakur \\
Physics Department, Alipurduar College, Dist. : Jalpaiguri, India
\\
A. Saha \\
Physics Department, Darjeeling Government College, Darjeeling, India}

\date{}

\maketitle

\vspace{0.5in}

\begin{abstract}

We present a holographic dark energy model of the universe
considering modified  generalized Chaplygin gas (GCG). The modified
GCG behaves as an ordinary barotropic fluid in the early epoch when
the universe was tiny but behaves subsequently as a $\Lambda$CDM
model at late epoch. An equivalent model with scalar field is
obtained here by constructing the corresponding potential. The
holographic dark energy is identified with the modified GCG and we
determine the corresponding holographic dark energy field and its
potential. The stability of the holographic dark energy in this case
is also discussed.

\end{abstract}

\vspace{0.2cm}

PACS number(s) : 04.50.+h, 98.80.Cq

\vspace{4.5cm}

\pagebreak

\section{ Introduction:}

In the recent years a number of observational facts from high
redshift surveys of type Ia Supernovae, WMAP, CMB etc. led us to
believe that our universe is passing through an accelerating phase
of expansion [1]. It is  generally accepted that our universe might
have also emerged from an accelerating phase in the past.  Thus
there might have two phases of acceleration of the universe (i)
early inflation and (ii) late acceleration followed by a
decelerating phase. It is known that perfect fluid assumption in the
framework of Einstein general theory of Relativity (GTR)  cannot
fully explain the observational facts in the universe.  It is known
that  early inflation may be realized in a  semiclassical theory of
gravity where matter is described by quantum fields [2]. Starobinsky
obtained inflationary  solution considering a curvature squared term
in the Einstein-Hilbert action [3] long before the advent of
inflation was known. However, the efficacy of the model is known
only after the seminal work of Guth who first employed the phase
transition mechanism to accommodate inflation. Thus inflation may be
realized either modifying the matter sector or the gravitational
sector of the Einstein's field equation. A number of literature
appeared in the last few years in which curvature squared terms [4]
are added to the Einstein-Hilbert action to build early inflationary
universe scenario.

Alternatively modification in the matter sector with an equation of
state $p = \omega \; \rho$, permits inflation in the early universe
if $\omega = -1$ but for the present acceleration one requires
$\omega < -1$. The usual fields in the standard model of particle
physics are not suitable to obtain the late accelerating phase of
the universe. Thus it is a challenge to theoretical physics to
formulate a framework to accommodate the observational facts. In
order to describe the present accelerating phase of the universe, it
may be useful to consider dark energy in the theory. The
observational facts in the universe predict that dark energy content
of the universe is about 76 \% of the total energy budget of the
universe.
 To accommodate such a huge energy  various kinds of exotic matters in the theory
 are considered to identify possible candidate for the dark energy.
  Chaplygin gas is
considered to be one such candidate  of dark energy
 with  an equation of state $p = - \frac{B}{\rho}$ [5], where $\rho$ and $p$
 are the energy density and
 pressure respectively and $B$ is a constant. Subsequently, a modified  form of
 the equation of state
 $p = - \frac{B}{\rho^{\alpha}}$ with $0 \leq \alpha \leq 1$
 was considered to construct a viable
 cosmological model [6, 7],    which is  known as generalized
 Chaplygin gas (GCG) in cosmology. It has two free parameters.
 It behaves initially like  dust but subsequently
  evolves to an asymptotic cosmological constant at late time  when the
  universe is sufficiently large.
 The  GCG behaves as
   a fluid obeying an equation of state  $p = \omega \rho$ at a later epoch. To
   accommodate dark energy various kinds of matter with a modified equation
   of state are also
 considered in the literature. However, recently another form of equation
 of state for Chaplygin gas [8] is proposed similar to that
 considered in [9], which is given by
 \begin{equation}
p = A \rho - \frac{B}{\rho^{\alpha}} \; \;  \; with \; \; 0 \leq
\alpha \leq 1,
\end{equation}
where $A$ is an equation of state parameter and $B $ is a constant,
known as modified GCG. This has three free parameters. In the early
universe when the size of the universe $a(t)$ was small, the
modified GCG gas corresponds to a barotropic fluid (if one considers
$A = \frac{1}{3} $ it corresponds to radiation and $A=0$ it
corresponds to matter). So, the modified GCG at one extreme end
behaves as an ordinary fluid and at the other extreme when the
universe is sufficiently large it behaves as cosmological constant
which can be fitted to  a $\Lambda$CDM model. In a flat Friedmann
model it is shown [6]  that the modified generalized Chaplygin gas
may be equivalently described in terms of
 a homogeneous minimally coupled scalar field $\phi$. Barrow [10] has outlined a method  to fit
 Chaplygin gas in FRW universe. Gorini {\it  et al.} [11] using the above scheme obtained
 the corresponding
homogeneous scalar field $\phi (t)$  in a  potential $V(\phi)$ which
can be used to obtain a viable cosmological model with modified
Chaplygin gas.

Recently, holographic principle [12, 13] is incorporated in
cosmology [14-17] to track the dark energy content of the universe
following the work of Cohen {\it et al.} [18]. Holographic principle
is a
 speculative conjecture about quantum gravity theories proposed by G't Hooft. The idea is subsequently
 promoted by Fischler and Susskind [12] claiming that all the information contained in a spatial volume may
  be represented by a theory that lives on the boundary of that space.
For a given finite region of space it may  contain matter and energy
within it. If this energy  suppresses a critical   density then the
region collapses to a black hole. A black hole is known
theoretically to have an entropy which is proportional to its
surface area of its event horizon. A black hole event horizon
encloses a volume, thus a  more massive  black hole have larger
event horizon and encloses larger volume. The most massive black
hole that can fit in a given region is the one whose event horizon
corresponds exactly to the boundary of the given region under
consideration. The maximal limit of entropy for an ordinary region
of space is directly proportional to the surface area of the region
and not to its volume. Thus, according to holographic principle,
under  suitable conditions all the information about a physical
system inside a spatial region is encoded in the boundary. The basic
idea of a holographic dark energy in cosmology is that
 the saturation of the entropy bound may be related to an  unknown ultra-violet (UV)  scale
$\Lambda$ to some known comological scale
  in order to enable it to find a viable formula for the dark energy which may be quantum gravity in
  origin and it is  characterized by $\Lambda$. The choice of UV-Infra Red (IR) connection from the covariant
  entropy bound leads to a universe dominated by blackhole states.
According to   Cohen {\it et al.} [18] for   any state in the
Hilbert space with energy $E$, the  corresponding  Schwarzschild
radius $R_s \sim E$, may be less than the IR cut off value $L$
(where $L$ is a cosmological scale). It is possible to derive a
relation between the UV  cutoff $\rho_{\Lambda}^{1/4}$ and the IR
cutoff which eventually
   leads to a constraint   $ \left( \frac{8 \pi G}{c^2} \right) L^3  \left( \frac{\rho_{\Lambda}}{3}
   \right) \leq L$ [19] where $\rho_{\Lambda}$ is the energy density corresponding to dark energy
   characterized by $  \Lambda$, $G$ is Newton's gravitational constant  and $c$ is a parameter in the theory.
   The holographic dark energy density  is
\begin{equation}
\rho_{\Lambda} = 3 c^2 M_P^2 L^{-2},
\end{equation}
where $M_P^{-2} = 8 \pi G$.
It is known that the present acceleration may be described if $\omega_{\Lambda} =\frac{p_{\Lambda}}{\rho_{\Lambda}} < - \frac{1}{3}$. If one considers $L \sim \frac{1}{H}$ it gives  $\omega_{\Lambda} =0$.
A holographic cosmological constant model based on Hubble scale as IR cut off does not permit accelerating universe.
It is also  examined [14]  that the holographic dark energy model based on  the particle horizon as the IR cutoff even does not work to get an accelerating universe. However, an alternative model of dark energy  using particle horizon in closed model is also proposed [20].
However, Li [15] has obtained an accelerating universe considering  event horizon  as the cosmological scale. The model is consistent with the cosmological observations.  Thus to have a model consistent with observed universe one should adopt the covaiant entropy bound and choose $L$ to be event horizon.

The paper is organized as follows : in sec. 2, the relevant field
equation with modified Chaplygin gas in FRW universe is presented;
in sec. 3 we present an equivalent model with a scalar field by
constructing the corresponding potential, in sec. 4, holographic
dark energy fields in modified GCG is determined;  in sec. 5,
squared speed of sound for holographic dark energy is evaluated for
a closed universe i.e., $k = 1$. Finally in sec. 6,  a brief
discussion.

\section{ Modified Chaplygin Gas in FRW universe :}

The Einstein's field equation is
given by
\begin{equation}
G_{\mu \nu} =  \kappa^2 \; T_{\mu \nu}
\end{equation}
where $\kappa^2 = 8 \pi G$,  and $T_{\mu \nu}$ is the energy
momentum tensor.

We consider a homogeneous and isotropic universe given  by
\begin{equation}
ds^{2} = - dt^{2} + a^{2}(t) \left[ \frac{dr^{2}}{1- k r^2} + r^2 ( d\theta^{2} + sin^{2} \theta \;
d  \phi^{2} ) \right]
\end{equation}
where $a(t)$ is the scale factor of the universe, the matter is
described by the energy momentum tensor $T^{\mu}_{\nu} = ( \rho, p,
p, p)$ where  $\rho$ and $p$ are energy density and pressure
respectively.

Using the metric (4) and  the energy momentum tensor, the  Einstein's field equation (3)  can be written as
\begin{equation}
H^{2}+\frac{k}{a^2} =  \frac{1}{ 3 M_{P}^2 } \rho
\end{equation}
where we use $M_{P}^{2} = \kappa^2$. The conservation equation for
matter is given by
\begin{equation}
\frac{d\rho}{dt} + 3 H (\rho + p) = 0 .
\end{equation}
For   modified generalized Chaplygin gas given by eq. (1), the
energy density is obtained from eq. (6), which is given by
\begin{equation}
\rho = \left(  \frac{B}{1+A} + \frac{C}{a^n} \right)^{\frac{1}{1+
\alpha}}
\end{equation}
where $B$ and $C$ are arbitrary integration constants and we denote
$ n = 3(1 + A) ( 1 + \alpha)$ . We now define the following
\begin{equation}
\Omega_{\Lambda} = \frac{\rho_{\Lambda}}{\rho_{cr}}, \; \Omega_{m} = \frac{\rho_{m}}{\rho_{cr}}, \;
\Omega_{k} = \frac{k}{a^2 H^2}
\end{equation}
where $\rho_{cr} = 3 M_P^2 H^2$, $\Omega_{\Lambda}$, $ \Omega_m $
and $\Omega_k$ represent  density parameter corresponding to
$\Lambda$, matter and curvature respectively.

\section{Modified GCG as a scalar field :}

We assume  here that the origin of dark energy is a scalar field. In
this case, using Barrow's scheme [10], we get the following
\begin{equation}
\rho_{\phi} = \frac{1}{2} \dot{\phi}^2 + V (\phi) = \left(
\frac{B}{A+1} +  \frac{C}{a^n} \right)^{\frac{1}{\alpha +1}},
\end{equation}
\begin{equation}
p_{\phi} = \frac{1}{2} \dot{\phi}^2 - V (\phi) = \frac{ - \frac{
B}{A + 1} +  A \frac{C}{a^n}}{
 \left( \frac{ B}{A+1} +  \frac{C}{a^n} \right)^{\frac{\alpha}{\alpha
 +1}}}.
\end{equation}
It is now simple to derive the scalar field potential and its
kinetic energy  term, which are given by
\begin{equation}
V(\phi) = \frac{  \frac{B}{A + 1} + \frac{1 - A}{2}  \frac{B}{
a^n}}{ \left( \frac{ B}{A+1} +  \frac{C}{a^n}
\right)^{\frac{\alpha}{\alpha +1}}},
\end{equation}
\begin{equation}
\dot{\phi}^2 = \frac{ (A + 1) \frac{ C}{ a^n}}{  \left( \frac{
B}{A+1} +  \frac{C}{a^n} \right)^{\frac{\alpha}{\alpha +1}}}.
\end{equation}

We now describe two cases :

Case I :  For a flat universe ($k = 0$),  using eq. (5) one can
integrate eq. (12) which yields
\begin{equation}
\phi - \phi_o = \pm \frac{2 M_P}{\sqrt{n (1 + \alpha)}} \; sinh^{-1}
\left[ \sqrt{ \frac{C (A + 1)}{B}} a^{- \frac{n}{2}} \right]
\end{equation}
and the  corresponding potential is given by
\begin{equation}
V(\phi) = \frac{ \frac{B}{1 + A} + \frac {B (1 - A) }{2 (1 + A)} \;
sinh^2 \left( \frac{ \sqrt{n}}{2} \; (\phi - \phi_o) \right) }{
\left( \frac{B}{1 + A} \right)^{\frac{\alpha}{\alpha + 1}}  \;
cosh^{\frac{2 \alpha}{1 + \alpha}}  \left( \frac{ \sqrt{n}}{2} \;
(\phi - \phi_o) \right)}.
\end{equation}

The potential behaves as a constant near $\phi \rightarrow \phi_o$,
otherwise it increases with increasing value of the field.

Case II : We now consider a non flat universe, in this case the
evolution of the scalar field is obtained as
\begin{equation}
\phi - \phi_o = \pm \int \sqrt{ \frac{12 M_P^{2} (A + 1)}{n^2}} \;
\frac{dz}{\left( \mu^2+z^2 \right)  -\frac{3 M_P^{2} K}{C^{2/n}} \;
z^{4/n} (\mu^2+z^2)^{\frac{\alpha}{\alpha +1}}}
\end{equation}
where $k = +1$ for closed universe ( $k = -1$ for open universe )
and we denote $\mu^2= \frac{B}{A+1}$, $z= \sqrt{\frac{c}{a^n}}$. The
integral may be evaluated analytically for some special choice of
the parameters. We choose the following :

$\bullet$ $A = - \frac{1}{3}, \; \; B = 0$, the scalar field evolves
as
\begin{equation}
\phi_{\pm}  = \phi_o \pm \sqrt{ \frac{ 2 M_P^2}{n^2(1 - \frac{3
M_P^2 k}{c^{2/n} })}} \; ln \; \left( \frac{C}{a^n} \right),
\end{equation}
the corresponding scalar field potential is given by
\begin{equation}
V(\phi) = \frac{2}{3} \left( \frac{2 M_P^2}{n^2(1- \frac{3M_P^2
k}{c^{2/n}} )}^{\frac{1}{2(\alpha+1)}}  \right) \; e^{\pm \;
\frac{1}{\alpha +1} \; (\phi - \phi_o)}.
\end{equation}
The potential is exponential in nature, it is increasing or
decreasing depending on the type of inflaton field one choose in
this case. We also note that  a positive  potential is obtained for
$C
> \left( 3 M_P^2 k \right)^{n/2}$. In the case of closed universe
the above inequality gives lower bound on the value of $C$. But in
the case of an open universe $C$ can pick up  negative values also
for even integer values of $n$.

 $\bullet$ $A = \frac{1}{3}, \; \; B \neq 0$, in
this case the scalar field evolves as
\begin{equation}
\phi - \phi_o = \pm \sqrt{ \frac{ 16 M_P^2}{n^2(1 - \frac{3 M_P^2
k}{c^{2/n}})}} \; sinh^{-1}\; \left(\frac{4C}{3B} \frac{1}{a^{n/2}}
\right),
\end{equation}
and the corresponding potential is given by
\begin{equation}
V(\phi) = sech^2 \; \left( \frac{n^2(1- \frac{3M_P^2 k}{c^{2/n}}
)}{16 M_P^2} \right)  \; (\phi - \phi_o)  + \frac{1}{3} \; tanh^2
\;\left( \frac{n^2(1- \frac{3M_P^2 k}{c^{2/n}} )}{16 M_P^2} \right)
\; ( \phi - \phi_o).
\end{equation}
The potential is drawn in fig. 1. It is a new and interesting
potential, it has a shape similar to that one obtains in the case of
tachyonic field [21]. The difference is that there is an extra term
to the potential which attains a constant value for large value of
the inflaton field. This potential may be important for a viable
cosmological model building. We also note that it permits an
oscillatory scalar field for $C < \left( 3 M_P^2 k \right)^{n/2}$
with sinusoidal potential for any $n$ in closed universe but for
open universe an even integer value of $n$ only gives such
potential.

\input{epsf}
\begin{figure}
\epsffile{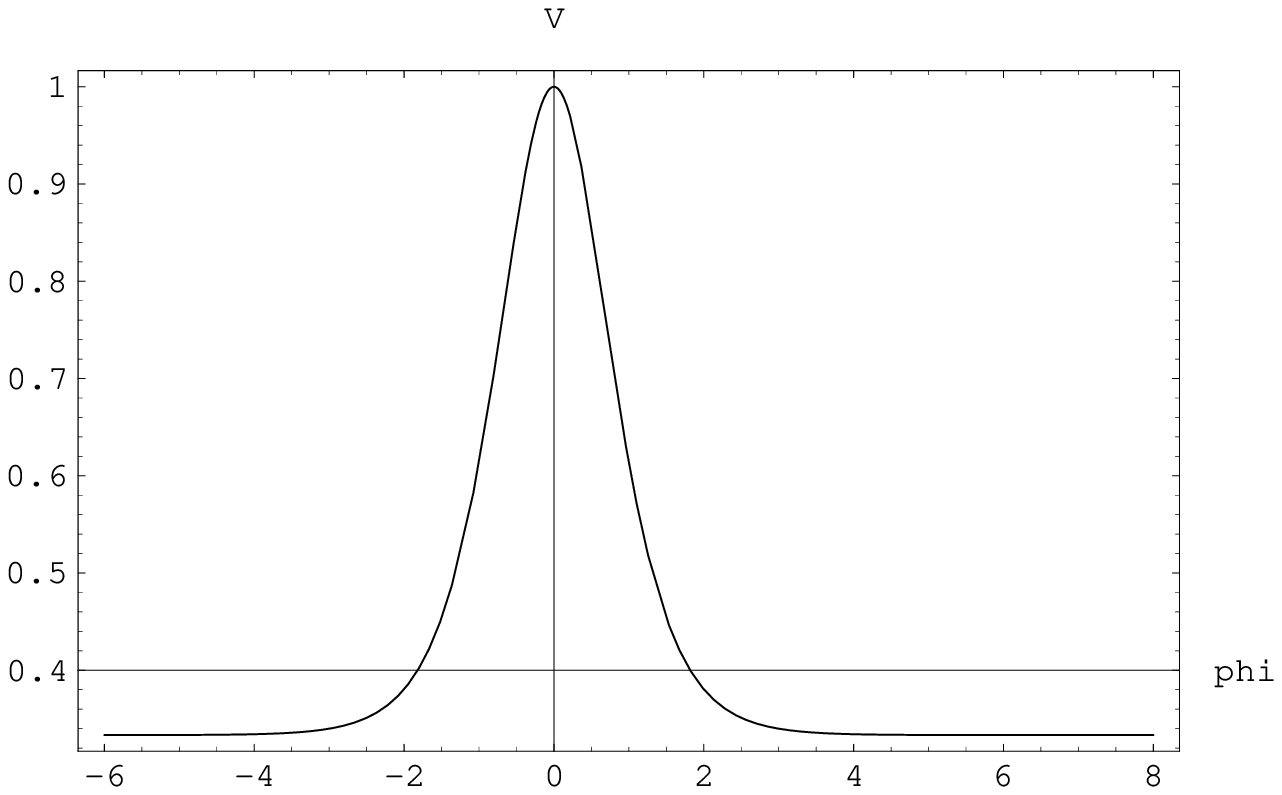} \caption{shows the plot of $V$ versus $\phi$
with the parameter $ \frac{n^2(1- \frac{3M_P^2 k}{c^{2/n}} )}{16
M_P^2} =1$.}
\end{figure}

\section{Holographic Dark Energy in Modified GCG :}

In a FRW universe we now consider a non-flat universe with $k \neq 0$ and use the holographic dark energy density as given in (2) which is
\begin{equation}
\rho_{\Lambda} = 3 c^2 M_P^2 L^{-2},
\end{equation}
where $L$ is the cosmological length scale for tracking the field
corresponding to holographic dark energy in the universe. The
parameter $L$ is defined as
\begin{equation}
L = a r (t).
\end{equation}
where $a(t)$ is the scale factor of the universe and $r(t)$ is relevant to the future event horizon of the universe. Using Robertson-Walker metric one gets [16]
\[
L = \frac{a (t)}{\sqrt{|k|}}  \;  sin \; \left[ \sqrt{|k|}
R_{h}(t)/a(t) \right] \;\; for \; \; \; k = +1 ,
\]
\[
\; \; \; =   R_h \;\; for \; \; k =0,
\]
\begin{equation}
 \; \; \; =  \frac{a (t)}{\sqrt{|k|}} \; \; sinh \; \left[
\sqrt{|k|} R_{h}(t)/a(t) \right]\; \; for \; \; k = - 1 .
\end{equation}
where $R_{h}$ represents the event horizon which is given by
\begin{equation}
R_h = a(t) \; \int_t^\infty \frac{dt'}{a(t')} = a(t) \; \int^{r_1}_o \frac{dr}{\sqrt{1 - k r^2}}.
\end{equation}
 Here $R_h$ is
 measured in $r$ direction and $L$ represents the radius of the event horizon measured on the sphere of the horizon. Using the definition of $\Omega_{\Lambda} = \frac{\rho_{\Lambda}}{\rho_{cr}} $ and $\rho_{cr} = 3 M_{P}^2 H^2$, one can derive  [17]
\begin{equation}
H L = \frac{c}{\sqrt{\Omega_{\Lambda}}}.
\end{equation}
Using eqs. (17)- (18), we determine the rate of change of $L$ with respect to
$ t$  which is
\[
\dot{ L} = \frac{c}{\sqrt{\Omega_{\Lambda}}} - \frac{1}{\sqrt{|k|}}
\; cos \; \left( \frac{\sqrt{|k|} \; R_h}{a(t)} \right) \; \; for \;
\; k = +1,
\]
\[
\; \; = \frac{c}{\sqrt{\Omega_{\Lambda}}} - 1 \; \; for \; \; k = 0,
\]
\begin{equation}
\; \; \; = \frac{c}{\sqrt{\Omega_{\Lambda}}} - \frac{1}{\sqrt{|k|}}
\; cosh \; \left( \frac{\sqrt{|k|} \; R_h}{a(t)} \right) \; \; for
\; \; k = - 1.
\end{equation}
Using eqs. (15) -(20) , it is possible to construct the required equation for
the holographic energy density $\rho_{\Lambda}$, which is given by
\begin{equation}
\frac{d\rho_{\Lambda}}{dt} =  - 2 H \left[ 1 -
\frac{\sqrt{\Omega_{\Lambda}}}{c} \; \frac{1}{\sqrt{|k|}} \; f(x)
\right]
 \; \rho_{\Lambda},
\end{equation}
where we use the notation, henceforth,
\begin{equation}
 f(X) = \frac{1}{\sqrt{|k|}} \; cosn \left( \sqrt{|k|} \; x \right) =
 cos (X) \; \left[ 1, cosh (X) \right]  \; fo r\;   k =1 \; [0, -1],
\end{equation}
where $X = \frac{R_h}{a(t)}$.
 The energy conservation equation is
\begin{equation}
\frac{d\rho_{\Lambda}}{dt} + 3   H (1 + \omega_{\Lambda})  \rho_{\Lambda} = 0
\end{equation}
which is used to determine the the equation of state parameter
\begin{equation}
\omega_{\Lambda}  =  - \left( \frac{1}{3} +  \frac{2
\sqrt{\Omega_{\Lambda}}}{3c} f(X) \right).
\end{equation}
Now  we assume  holographic dark energy density which is equivalent to  the modified  Chaplygin gas energy density.  The corresponding  energy density may be obtained  using the equation of state given by  (7).
The equation of state parameter using (1) can also be re-written as
\begin{equation}
\omega  = \frac{p}{\rho} =  A  -   \frac{B}{\rho^{\alpha +1}}.
\end{equation}
Let us now establish the correspondence between the holographic dark
energy and modified Chaplygin gas energy density. In this case from
eqs. (7) and (20), we get
\begin{equation}
C = a^n \left[(3 c^2 M_P^2 L^{-2})^{1+\alpha} - \frac{B}{A+1}
\right].
\end{equation}
Thus using eqs.  (29)-(30) in the above we determine the parameters
as
\begin{equation}
B = ( 3 c^2 M_P^2 L^{-2}) ^{\alpha +1} \; \left[ A + \frac{1}{3} +
\frac{2  \sqrt{\Omega_{\Lambda}}}{3  c}  f(X) \right],
\end{equation}
\begin{equation}
C = ( 3 c^2 M_P^2 L^{-2})^{\alpha +1} \; a^n \left[ 1 - \frac{3 A
+1}{3(A + 1)} - \frac{2 \sqrt{\Omega_{\Lambda}}}{3 (A + 1) c} f(X)
\right].
\end{equation}
 The scalar field potential  becomes
\begin{equation}
V( \phi)  = 2  c^2 M_P^2 L^{-2} \left[ 1 +  \frac{
\sqrt{\Omega_{\Lambda}}}{2 c} f(X) \right],
\end{equation}
and the corresponding kinetic energy of the field is given by
\begin{equation}
\dot{ \phi}^2  = 2  c^2 M_P^2 L^{-2} \left[ 1 -  \frac{
\sqrt{\Omega_{ \Lambda}}}{  c} f(X) \right].
\end{equation}
Considering $x$ $(= \ln a)$, we transform  the time derivative to the derivative with logarithm of the  scale factor,  which is the most useful function in this case. We get
\begin{equation}
\phi'= M_P \sqrt{ 2 \Omega_{\Lambda} \left( 1 - \frac{
\sqrt{\Omega_{\Lambda}}}{  c} f(X) \right)}
\end{equation}
where $()'$ prime represents derivative with respect to $x$. Thus, the evolution of the calar field is given by
\begin{equation}
\phi (a) - \phi ( a_o) = \sqrt{2} M_P \int_{\ln a_o}^{\ln a} \sqrt{
\Omega_{\Lambda} \left( 1 -  \frac{ \sqrt{\Omega_{\Lambda}}}{  c}
f(X) \right)} \;  dx  .
\end{equation}

\section{Squared speed for  Holographic Dark Energy :}

We consider a closed  universe model ($k = 1$) in this case. The
dark energy equation of state parameter given by eq. (29) reduces to
\begin{equation}
\omega_{\Lambda}  = - \frac{1}{3} \left( 1 + \frac{2}{c}
\sqrt{\Omega_{\Lambda}} \; cos \; y \right)
\end{equation}
where  $y = \frac{R_H}{a(t)}$. The minimum value it can take is
 $\omega_{min} = - \frac{1}{3}  \left( 1 + 2 \sqrt{\Omega_{\Lambda}} \right)$ and one obtains
 a lower bound $\omega_{min} = - 0.9154$ for
 $\Omega_{\Lambda}= 0.76$ with $c = 1$.
Taking variation of the state parameter with respect to $x = \ln \; a$, we get [17]
\begin{equation}
\frac{\Omega_{\Lambda}'}{\Omega_{\Lambda}^2}= (1 - \Omega_{\Lambda})
\left( \frac{2}{c} \frac{1}{\Omega_{\Lambda}} cos \; y + \frac{1}{1
- a \gamma} \frac{1}{\Omega_{\Lambda}} \right)
\end{equation}
and the variation of equation of state parameter becomes
\begin{equation}
\omega_{\Lambda}' = - \frac{\sqrt{\Omega_{\Lambda}}}{3 c} \left[
\frac{1 - \Omega_{\Lambda}}{1 - \gamma a} + \frac{ 2
\sqrt{\Omega_{\Lambda}}}{c} \left(1 - \Omega_{\Lambda} cos^2
y\right) \right],
\end{equation}
where $\gamma = \frac{\Omega^{o}_k}{\Omega^{o}_m}$. We now introduce
the squared speed of holographic dark energy fluid  as
\begin{equation}
{\it v}_{\Lambda}^2 = \frac{dp_{\Lambda}}{d \rho_{\Lambda}} = \frac{\dot{p}_{\Lambda}}{\dot{\rho}_{\Lambda}} = \frac{p'_{\Lambda}}{\rho'_{\Lambda}},
\end{equation}
where  varaiation of eq. (30) w.r.t.  $x$ is  given by
\begin{equation}
p'_{\Lambda} = \omega'_{\Lambda} \rho_{\Lambda}+ \omega_{\Lambda} \rho'_{\Lambda}.
\end{equation}
Using the eqs. (41) and (42) we get
\[
{\it v}_{\Lambda}^2 = \omega'_{\Lambda} \frac{{\rho}_{\Lambda}}{{\rho'}_{\Lambda}} + \omega_{\Lambda}
\]
which now becomes
\begin{equation}
{\it v}_{\Lambda}^2  = - \frac{1}{3}  - \frac{2}{3 c} \sqrt{\Omega_{\Lambda} } \; cos y + \frac{1}{6 c} \; \sqrt{ \Omega_{\Lambda} } \left[ \frac{ \frac{1 - \Omega_{\Lambda} }{1 - \gamma a}+ \frac{2}{c} \sqrt{ \Omega_{\Lambda} } \left( 1 - \Omega_{\Lambda} \; cos^2 y \right) }{ 1 - \frac{\Omega_{\Lambda} }{c} \; cos y  } \right].
\end{equation}

\input{epsf}
\begin{figure}
\epsffile{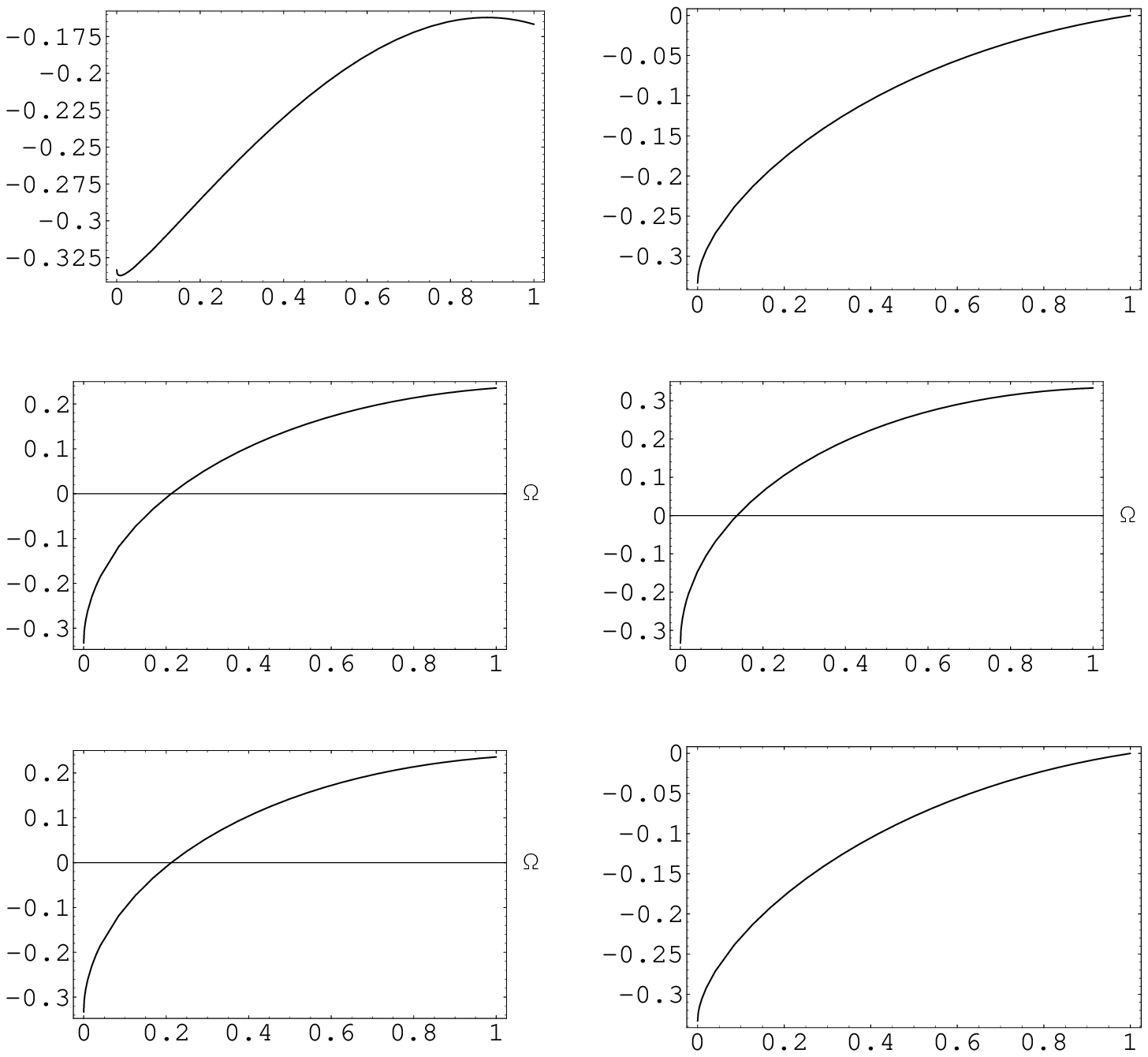}
\caption{shows the plot of ${\it v}_{\Lambda}^2$ versus $\Omega_{\Lambda} $ for different values of
$y$ with $c=1$, $\gamma = 1/3$ and $a=1$, in the first array the figures are  for $y =  \frac{\pi}{3} \; $ and $\; y =  \frac{\pi}{2} \; $, in the second array for  $ \; y =   \frac{1.5 \; \pi}{2} \; $,  $\; y =  \pi \; $ and in the third array for $y =   \frac{2.5 \; \pi}{2} \; $, $y =  \frac{3 \pi}{2}$.}
\end{figure}

The variation of ${\it v}_{\Lambda}^2$ with  $\Omega_{\Lambda}$ is shown in
 fig. 2 for different $y$ values. It is found that for a  given value of $c, \; a, \; \gamma$, the model admits a positive squared speed for $\Omega_{\Lambda} > 0$. However,
$\Omega_{\Lambda}$ is bounded below otherwise instability develops. We note also that for
 $  \frac{(2n+1) \pi}{2} < y < \frac{(2n+3) \pi}{2}$,
 (where $n$ is an integer)   no instability develops.
 We plot the case for $n = 0$ in  fig. 2, it is evident that for
 $y \leq \frac{\pi}{2}$ and $y \geq \frac{3 \pi}{2}$,  the squared speed
 for holographic dark energy  becomes negative which led to instability.
 But for the region $ \frac{\pi}{2} < y  <  \frac{3 \pi}{2}$ with $n=0$ no
 such instability develops. It is also found that for $y =0$ i.e., in flat case
 the holographic dark energy model is always unstable [22].

\section{  Discussions : }

In this paper we explored the holographic dark energy model in  FRW
universe with a scalar field which describes the modified
generalized Chaplygin gas (GCG).  We determine the equivalent scalar
field potential of the modified GCG in flat universe  which is
different from that obtained in {\it Ref.} 6. In the non-flat case
although it is not so simple to obtain an analytic function of the
potential in terms of the field we discuss two special cases in
which potentials are shown as a function of field $\phi$. For $A=
\frac{1}{3}, \; \; B \neq 0$, we obtain a new scalar field potential
similar to the potential of a rolling tachyon obtained earlier [21]
but with an extra piece in it.
 We also obtain the evolution of the holographic dark energy field
and the corresponding potential in the framework of modified GCG in
a non flat universe. Although recent observational evidence supports
a flat universe, it is not yet decided that our universe is
perfectly flat. Thus it is important to study  a closed or open
universe to account for the the observational facts. We recover the
equation of state considered by   Setare [21] when $A = 0$ and
$\alpha = 1$ to derive the fields of dark energy. We note that in
the closed model the holographic dark energy is stable for a
restricted domain of the values of $\Omega_{\Lambda}$. It is also
observed that inclusion of a barotropic fluid in addition to
Chaplygin gas (which is modified GCG) does not alter the form of
potential and evolution of the holograpic dark energy field but the
parameter $B$ in the equation of state becomes proportional to $a^n$
with $n = 3 (1 + A) (1 + \alpha)$. Thus  the contribution of the
holographic dark energy is more if ($ A \neq 0$) compared to the
case when one considers $A = 0$ as was done in {\it Ref.} [23]. Thus
it  is  noted that although the form of the potential does not have
impact on the addition of a barotropic fluid  it changes the overall
holographic dark energy density.  It is found that the
 holographic  dark energy  is stable for a restricted
 domain of the values of $\Omega_{\Lambda}$ in a closed model of the
  universe.

\vspace{0.5in}

{\large \it Acknowledgement :}

Authors would like to  thank IUCAA  Reference Centre at North Bengal
University (NBU) and Physics Department, NBU  for providing facility
to initiate the work. BCP would like to thank Third World Academy of
Sciences $\bf (TWAS)$ for awarding Associateship to visit {\it
Institute of Theoretical Physics, Chinese Academy of Sciences,
Beijing } and UGC for awarding Minor Research Project ({\it No. F.}
32-63/2006 (SR). BCP would like to thank {\it Miao Li } for a
fruitful discussion.

\pagebreak

\end{document}